\documentclass[reprint, amsmath, amssymb, aps, prl]{revtex4-2}

\usepackage{graphicx}
\usepackage{dcolumn}
\usepackage{bm}
\usepackage[backref=none]{hyperref} 
\usepackage[usenames,dvipsnames]{color}
\definecolor{MyDarkBlue}{rgb}{0,0.1,0.75}
\hypersetup{pdfborder={0 0 0},colorlinks,breaklinks=true,
  urlcolor={MyDarkBlue},citecolor={MyDarkBlue},linkcolor={MyDarkBlue}
}
\begin{document}

\preprint{APS/123-QED}

\title[Origin of ferroelectric domain wall alignment with surface trenches in ultrathin films]{Origin of ferroelectric domain wall alignment with surface trenches in ultrathin films}

\author{Jack S. Baker$^{1, 2}$}

\author{David R. Bowler$^{1, 2, 3}$}%

\affiliation{$^1$London Centre for Nanotechnology, UCL, 17-19 Gordon St, London WC1H 0AH, UK}

\affiliation{$^2$Department of Physics \& Astronomy, UCL, Gower St, London WC1E 6BT, UK}

\affiliation{$^3$International Centre for Materials Nanoarchitectonics (MANA) National Institute for Materials Science (NIMS), 1-1 Namiki, Tsukuba, Ibaraki 305-0044, Japan}

\date{\today}

\begin{abstract}
Engraving trenches on the surfaces of ultrathin ferroelectric (FE) films and superlattices promises control over the orientation and direction of FE domain walls (DWs). Through exploiting the phenomenon of DW-surface trench (ST) parallel alignment, systems where DWs are known for becoming electrical conductors could now become useful nanocircuits using only standard lithographical techniques. Despite this clear application, the microscopic mechanism responsible for the alignment phenomenon has remained elusive. Using ultrathin PbTiO$_3$ films as a model system, we explore this mechanism with large scale density functional theory simulations on as many as 5,136 atoms. Although we expect multiple contributing factors, we show that parallel DW-ST alignment can be well explained by this configuration giving rise to an arrangement of electric dipole moments which best restore polar continuity to the film. These moments preserve the polar texture of the pristine film, thus minimizing ST-induced depolarizing fields. Given the generality of this mechanism, we suggest that STs could be used to engineer other exotic polar textures in a variety of FE nanostructures as supported by the appearance of ST-induced polar cycloidal modulations in this letter. Our simulations also support experimental observations of ST-induced negative strains which have been suggested to play a role in the alignment mechanism.
\end{abstract}

\maketitle

Ferroelectric (FE) domain walls (DWs) have been observed in parallel alignment with surface and substrate defects since the 2000's \cite{Streiffer2002, Fong2004, Thompson2008, Hadjimichael2018, Park2018}. While this phenomenon was initially proposed to control polarization dependent surface reactions \cite{Thompson2008}, now that electrically conducting FE DWs \cite{Seidel2009, Hong2021} have been realized, new horizons for nanoelectronic devices approach. That is, the phenomenon now promises control over the orientation of 2D conducting channels on the nanoscale; a potential pathway for fabricating DW-mediated nanocircuits with standard lithographical techniques. Presently, control over FE DWs and polar textures is achieved using carefully directed electric fields. With this technique, exotic topological phases including polar skyrmions \cite{Das2020} have been stabilized and DW-based nanocomponents \cite{Whyte2015, McConville2020, Sharma2017, Jiang2017} have been created; the latter defining the emerging field of \textit{DW nanoelectronics} via \textit{DW injection} \cite{Maksymovych2011, Catalan2012, Whyte2013, Whyte2015, Werner2017}. While directed electric fields offer the advantages of nanocircuits written with reversibility \cite{Hong2021}, they can lack permanence once the fields are released. Defect mediated alignment, however, offers enduring control without external stimuli. In light of this, we envisage great opportunities for these techniques to be applied in tandem helping to enable the next generation of DW-nanoelectronic devices. \par

Notwithstanding these advances, the mechanism for DW alignment with surface defects is far from a settled topic. One study used macroscopic time-dependent Landau–Ginzburg–Devonshire theory to examine the interaction of PbTiO$_3$ (PTO)/SrTiO$_3$ (STO) superlattices with the milled edges of the nanostructure \cite{Park2018}. They suggested that alignment arises from lowering the bulk and electrostrictive energy terms driven by the release of lateral mechanical restraints near the edge. Another study used a nanofocused X-ray beam to reveal giant strain and strain gradients near milled surface defects on the PTO/STO superlattice \cite{Hadjimichael2018}. This opens the door to the possibility of piezoelectric and/or flexoelectric contributions to the alignment mechanism. Until the present letter, no microscopic origin for the phenomenon has been proposed. \par

The feature of the domain structure pinned in the alignment is also disputed. A density functional theory (DFT) study of steps on the PTO (001) surface \cite{Shimada2010} found that DWs formed exactly at the step edges, suggesting that the DWs themselves are pinned. In contrast, an effective Hamiltonian study \cite{Prosandeev2007} found that film stability was enhanced when the domain centroid (DC, the area of maximal out-of-plane polarization furthest from the DW) was pinned by the step edge. While the DFT study \cite{Shimada2010}, in principle, should offer higher accuracy predictions, like-for-like comparisons cannot be made since the effective Hamiltonian study \cite{Prosandeev2007} treats the underlying flux-closure domain structure \cite{Tang2015, Baker2020Polar} when the former does not. Indeed, providing a full treatment of FE domains with DFT quickly becomes  intractable due to the $\mathcal{O}(N^3)$ (where $N$ is the number of atoms) scaling wall for computational time present in conventional DFT codes \cite{Bowler2012, Goedecker1999}. \par
\begin{figure}
       \centering
       \includegraphics[width=\linewidth]{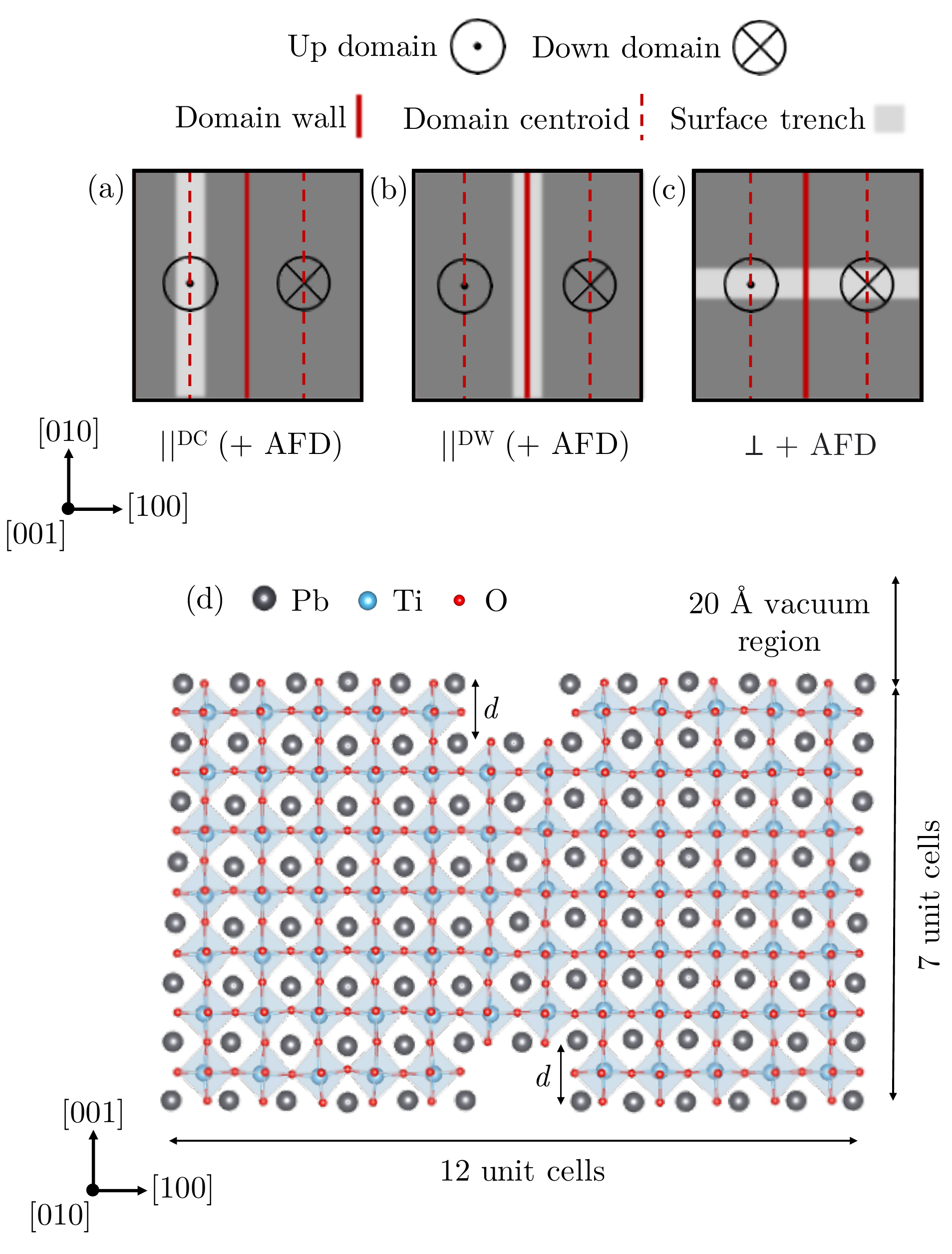}
        \caption{(a-c) Bird's eye views of ST configurations. +AFD is bracketed in (a) and (b) as they are treated with and without AFD modes. (a) Parallel to a domain wall, positioned over a domain wall: $\parallel^{\text{DW}}$. (b) Parallel to a domain wall, positioned over a domain centroid: $\parallel^{\text{DC}}$. (c) Perpendicular to a domain wall: $\perp$ + AFD. d) Looking down the axis (the [010] direction) of the $d=1$ $\parallel^{\text{DW}}$ film.}
        \label{fig:TrenchConfigs}
\end{figure}
In this letter, we address these discrepancies and offer a microscopic explanation of the DW alignment mechanism. Through performing state-of-the-art large scale DFT simulations of ultrathin PTO films, we venture far beyond previous works, able to treat the interaction between surface trenches (STs) and FE domains with DFT accuracy. We propose that DW-ST parallel alignment can be well explained by considering which ST orientations (relative to the DW) are best able to support continuity of the polar texture. This continuity minimizes new ST-induced depolarizing fields which would otherwise act to destabilize the polydomain polarization. We suggest that this alignment is only one consequence of a wider class of polar textures that can be produced through minimizing ST-induced depolarizing fields. This idea is reinforced by the appearance of engineered polar cycloidal modulations in our films. \par

To overcome the $\mathcal{O}(N^3)$ scaling wall of conventional DFT, the $\mathcal{O}(N)$ \cite{Goedecker1999, Bowler2012} scaling algorithm \cite{Li1993, Palser1998, Bowler1999} in the \textsc{CONQUEST} code \cite{Bowler2006, Nakata2020} (\texttt{v1.0.5} \cite{ZenodoDoi2020}) is used with a single-$\zeta$ plus polarization basis set of pseudoatomic orbitals \cite{Torralba2008, Bowler2019, Baker2020basis}, norm-conserving Hamann pseudopotentials \cite{Hamann2013, vanSetten2018} and the local density approximation as parameterized by Perdew \& Wang \cite{Perdew1992}. Other simulation details and preliminary tests are presented in the supplemental material \cite{supplement}. \par

We begin with a polydomain and free-standing PbO terminated PTO film seven unit cells in thickness, with a 20 $\text{\AA}$ vacuum region in the out-of-plane direction to prevent interactions between film images. An epitaxial strain of -$1.2\%$ (relative to the the in-plane lattice parameters of bulk $P4mm$ PTO) is imparted to represent the experimental PTO/STO lattice constant mismatch: the conditions where DW alignment has been experimentally observed \cite{Streiffer2002, Fong2004, Thompson2008, Hadjimichael2018, Park2018}. Explicit STO substrate is not used as the incipient broken inversion symmetry in the out-of-plane directions is known to give rise to sizeable perturbations to the domain structure \cite{Sai2000, Baker2020Polar} which we wish to isolate from our analysis. The equilibrium flux-closure domain period $\Lambda$ is used, which we find to be $12$ unit cells. One unit cell wide STs are placed on both surfaces of the film to (i) once again, limit the effect of inversion symmetry breaking and (ii) to eliminate spurious electric fields resultant from inequivalent surface dipole densities (otherwise requiring approximate correction \cite{Bengtsson1999}). STs are laterally separated by 12 unit cells in one of three positions: over the DC running parallel to the DW ($\parallel^{\text{DC}}$, Fig. \hyperref[fig:TrenchConfigs]{1(a)}), over the DW running parallel to the DW ($\parallel^{\text{DW}}$, Fig. \hyperref[fig:TrenchConfigs]{1(b)}) or running perpendicular to the DW (Fig. (\hyperref[fig:TrenchConfigs]{1(c)}). This last orientation features antiferrodistortive (AFD) modes since the even in-plane periodicity, broken surface symmetry and PbO termination will \textit{always} invoke the AFD $c(2 \times 2)$ surface reconstruction \cite{Munkholm2001} (enhanced surface $a^0a^0c^-$ octahedral rotations). We label these films $\perp$ + AFD. For fair comparison, we double the in-plane periodicity of $\parallel^{\text{DW/DC}}$ to additionally consider $\parallel^{\text{DW/DC}}$ + AFD films. We are then able to evaluate the impact of AFD modes on the DW alignment mechanism. \par

Each film listed so far is treated with ST depths of $d = 1, 2$ and $3$ unit cells with PbO terminated trench floors (shown in Fig. \hyperref[fig:TrenchConfigs]{1(d)}). To investigate the effects of lateral interactions between STs, we perform two further simulations of the $d=1$, $\parallel^{\text{DW/DC}}$ films with STs separated by $4 \Lambda$ ($\parallel^{\text{DW/DC}}_{4 \Lambda}$). This sufficiently limits the overlap of ST-induced surface strain fields. We relax each structure until the magnitude of the force on each atom falls below $0.01$ eV/$\text{\AA}$. \par

\begin{figure*}
       \centering
       \includegraphics[width=\textwidth]{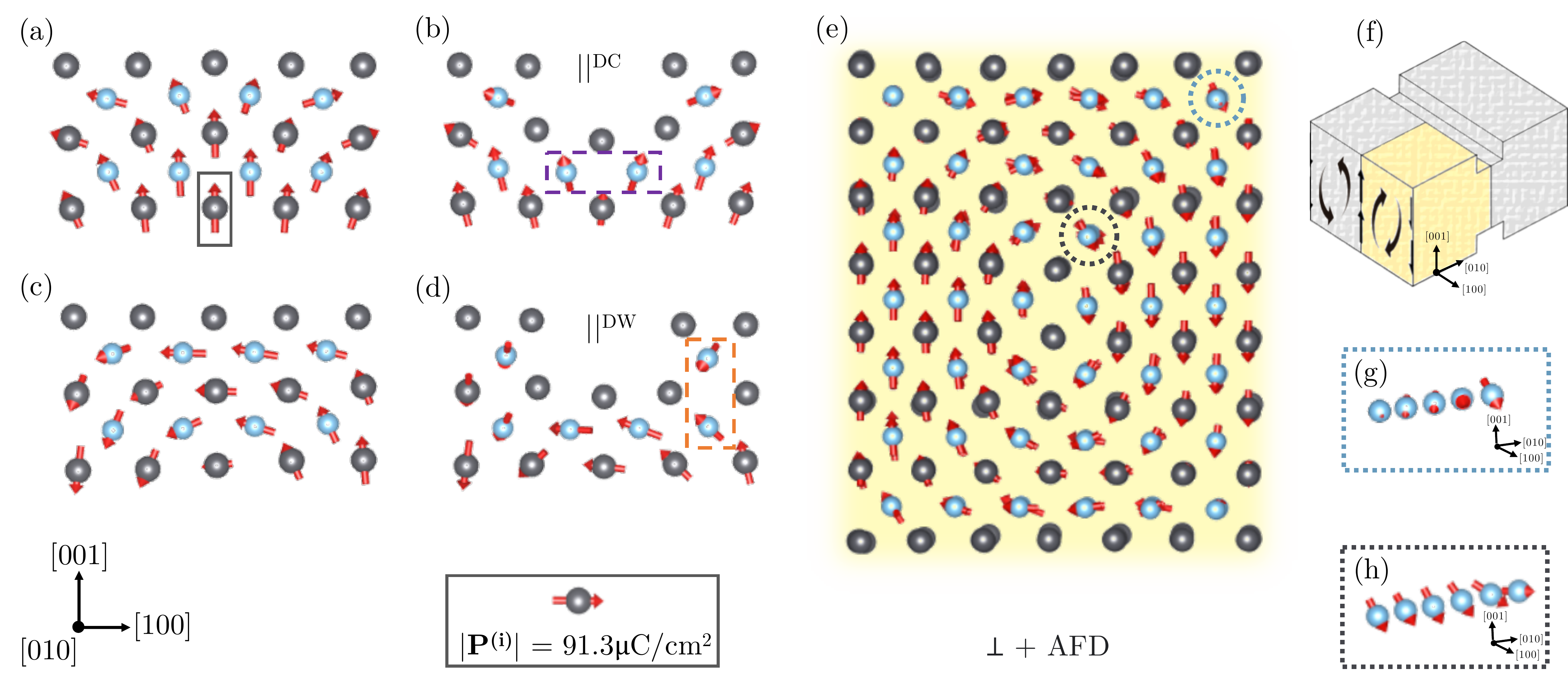}
        \caption{Local polarization vector fields calculated using the linear approximation first noted by Resta \cite{Resta1993}. Atom coloring is shared with Fig. \hyperref[fig:TrenchConfigs]{1} but O is removed for clarity. (a) At the DC of the pristine film. The Pb site in the grey box is used in the scale bar positioned below (d). (b) At the DC of the $d=1$ $\parallel^{\text{\text{DC}}}$ film. The purple dashed box is discussed in the text. (c) At the DW of the pristine film. (d) At the DW of the $d=1$ $\parallel^{\text{\text{DW}}}$ film. The orange dashed box is discussed in the text. (e) A single polar vortex of the $d=2$ $\perp$ + AFD film from the region colored on (f). (g) and (h) depict strings of Ti-centered local polarizations along the [010] direction from the color-matched dashed circles on (e).}
        \label{fig:VectorField}
\end{figure*}

The relative stability of each film is given in Table \hyperref[tab:TrenchEnergy]{I}. It is clear immediately that the most stable ST arrangement treated in this letter is $\parallel^{\text{DC}}$ + AFD, true for $d = 1, 2$ and $3$ supporting the results of an effective Hamiltonian study \cite{Prosandeev2007}. The hierarchy of stabilities is preserved independently of the treated $d$. In decreasing order we have $\parallel^{\text{\text{DC}}}$ + AFD, $\parallel^{\text{DW}}$ + AFD, $\perp$ + AFD and then the films without AFD modes: $\parallel^{\text{DC}}$ and $\parallel^{\text{DC}}$. This hierarchy demonstrates that parallel DW-ST alignment is favoured, in good agreement with experimental observations \cite{Streiffer2002, Fong2004, Thompson2008, Hadjimichael2018, Park2018}. For STs parallel to the DW, we see that part of this hierarchy is preserved independently of AFD modes. While this strongly implies that AFD modes play little-to-no role in the DW alignment mechanism, we do see that AFD modes greatly lower the total energy and lead to strong antiphase tilts (of $\approx 12^{\circ}$) at the surface which locally suppress polar modes in agreement with a study using a similar method \cite{Baker2020Polar}. Remarkably, these modes persist at a similar magnitude at the trench floor despite the broken connectivity with the surface octahedral tilt network. The $\parallel^{\text{DC}}_{4 \Lambda}$ film is also more stable than $\parallel^{\text{DW}}_{4 \Lambda}$ by 0.359 eV/1760 atoms; part of the hierarchy is unaffected by larger lateral ST separations, consistent with DW alignment observed for isolated defects \cite{Hadjimichael2018}. It is clear now that the DW alignment mechanism is robust to the conditions tested in this letter. \par

\begin{table}
\centering
\caption{Film stabilities relative to $\parallel^{\text{DC}}$ + AFD per film unit (FU): the number of atoms in the $\parallel$ arrangements. For $d = 1, 2$ and $3$, this is the energy per 428, 412 and 396 atoms, respectively.}
\label{tab:TrenchEnergy}
\resizebox{\linewidth}{!}{%
\begin{tabular}{@{}cccccc@{}}
\toprule 
& \multicolumn{5}{c}{E - E($\parallel^{\text{DC}}$ + AFD) {[}eV/FU{]}} \\
& & & & & \\
\multicolumn{1}{c|}{ }& \multicolumn{1}{c|}{$\parallel^{\text{DC}}$ + AFD} & \multicolumn{1}{c|}{$\parallel^{\text{DW}}$ + AFD} & \multicolumn{1}{c|}{$\perp$ + AFD} & \multicolumn{1}{c|}{$\parallel^{\text{DC}}$} & $\parallel^{\text{DW}}$ \\ \cline{1-6}
\multicolumn{1}{c|}{$d=1$} & \multicolumn{1}{c|}{0}        & \multicolumn{1}{c|}{+0.219}        & \multicolumn{1}{c|}{+1.706}        & \multicolumn{1}{c|}{+3.852}        & +4.106      \\
\multicolumn{1}{c|}{$d=2$} & \multicolumn{1}{c|}{0}        & \multicolumn{1}{c|}{+0.361}        & \multicolumn{1}{c|}{+0.881}        & \multicolumn{1}{c|}{+4.089}        & +4.095      \\
\multicolumn{1}{c|}{$d=3$} & \multicolumn{1}{c|}{0}        & \multicolumn{1}{c|}{+0.141}        & \multicolumn{1}{c|}{+1.951}        & \multicolumn{1}{c|}{+4.639}        & +4.878      \\ \botrule 
\end{tabular}
}
\end{table}

The stability of the different films can be best understood by considering how the polar texture adapts to a trench. That is, regardless of orientation, a trench must introduce new discontinuities to the polar texture of a pristine film (the film without STs) and therefore new depolarizing fields. The most stable arrangement must then be the one which \textit{best restores polar continuity}, thus minimizing ST-induced depolarizing fields. Using this principle, the favorability of $\parallel^{\text{DC}}$ over $\parallel^{\text{DW}}$ can be understood clearly from the local polarization vector fields shown in Fig. \hyperref[fig:VectorField]{2(a-d)}. Independent of depth, we find that the local polarization near the DC for the $\parallel^{\text{DC}}$ film (Fig. \hyperref[fig:VectorField]{2(b)}) is barely modulated in comparison to the DC of the pristine film (Fig. \hyperref[fig:VectorField]{2(a)}). This is made clear by the two Ti-centered polarization vectors in the purple dashed box of Fig. \hyperref[fig:VectorField]{2(b)} which reproduce the two surface modes present for the pristine film in the region for which the ST was inserted. This is because we have removed a unit cell at a site of preexisting out-of-plane polar discontinuity. The resulting polar texture arises from minimizing similar depolarizing fields to the pristine film. This contrasts greatly with the $\parallel^{\text{DW}}$ films (Fig. \hyperref[fig:VectorField]{2(d)}) where there are large differences in the local polarization compared with the pristine film (Fig. \hyperref[fig:VectorField]{2(c)}). In this case, the trench removes material in a region of continuous in-plane polarization at the cap of a polar vortex. This creates a new and large discontinuity in the polarization, giving rise to new in-plane depolarizing fields which are minimized by rotations of the local polarization at the trench edges towards [010] (or [0$\bar{1}$0]) and [00$\bar{1}$]. While this minimization increases the stability of the film, the resulting polar texture sees the two Ti-centered local polar modes in the orange box of Fig. \hyperref[fig:VectorField]{2(d)} stuck between a rock and a hard place. They are arranged in an electrostatically unfavorable near-head-to-head configuration. To compensate for this, the mean out-of-plane polarization of $\parallel^{\text{DW}}$ films are reduced by $\approx 5\%$ compared with $\parallel^{\text{DC}}$ films. All of these effects lower the stability of $\parallel^{\text{DW}}$ films. \par

The low stability of $\perp$ + AFD films can also be understood using the local polarization vector fields. This geometry is equivalent to interfacing an axial slice of the flux-closure domain structure of a thinner film with a thicker one. For example, at $d=2$, a single unit cell wide and 3 unit cell thick axial slice is inserted (STs are present on both surfaces). This slice has weaker out-of-plane polarization than the surrounding 7 unit cell thick film; an experimentally observed effect for thinner pristine films \cite{Fong2004}. In spite of this, the mean out-of-plane polarization for the entire film is greater than $\parallel^{\text{DW/DC}}$ + AFD films. The same is not true for the axial polarization (along [010], the Bloch components \cite{Shafer2018}). These become highly discontinuous giving rise to new depolarizing fields which are compensated for by strongly reducing the axial polarization. Discontinuity perpendicular to the direction of the polarization has striking effects. Fig. \hyperref[fig:VectorField]{2(e)} shows strong rotations of the local polarization in the (010) plane along the [010] direction within a single polar vortex (highlighted on Fig. \hyperref[fig:VectorField]{2(f)}). Fig. \hyperref[fig:VectorField]{2(g)} shows a sharp rotation as we approach the trench edge while Fig. \hyperref[fig:VectorField]{2(h)} shows a smooth rotation in order to create a continuous interface between the different flux-closure domains of the 7 and 3 unit cell thick regions. Although these modulations are cycloidal, close inspection reveals that the films remain achiral. However, degenerate chiral and achiral phases have been predicted in the PTO/STO system relating to the direction of the axial polarization for each polar vortex \cite{Shafer2018}. Since $\perp$ + AFD films feature suppressed axial polarization, these films actually suppress chirality. Further, these modulations give rise to inhomogeneous strain and strain gradients which interact with and disturb the underlying strain field of the pristine film. The prediction of trench-induced cycloidal modulations highlights that STs could be used to engineer exotic polar textures in ferroelectric nanostructures. All of these effects compounded with one another see the $\perp$ + AFD films most severely modulated from the pristine film affirming their low stability compared with $\parallel^{\text{DW/DC}}$ + AFD films. \par

\begin{figure}
       \centering
       \includegraphics[width=\linewidth]{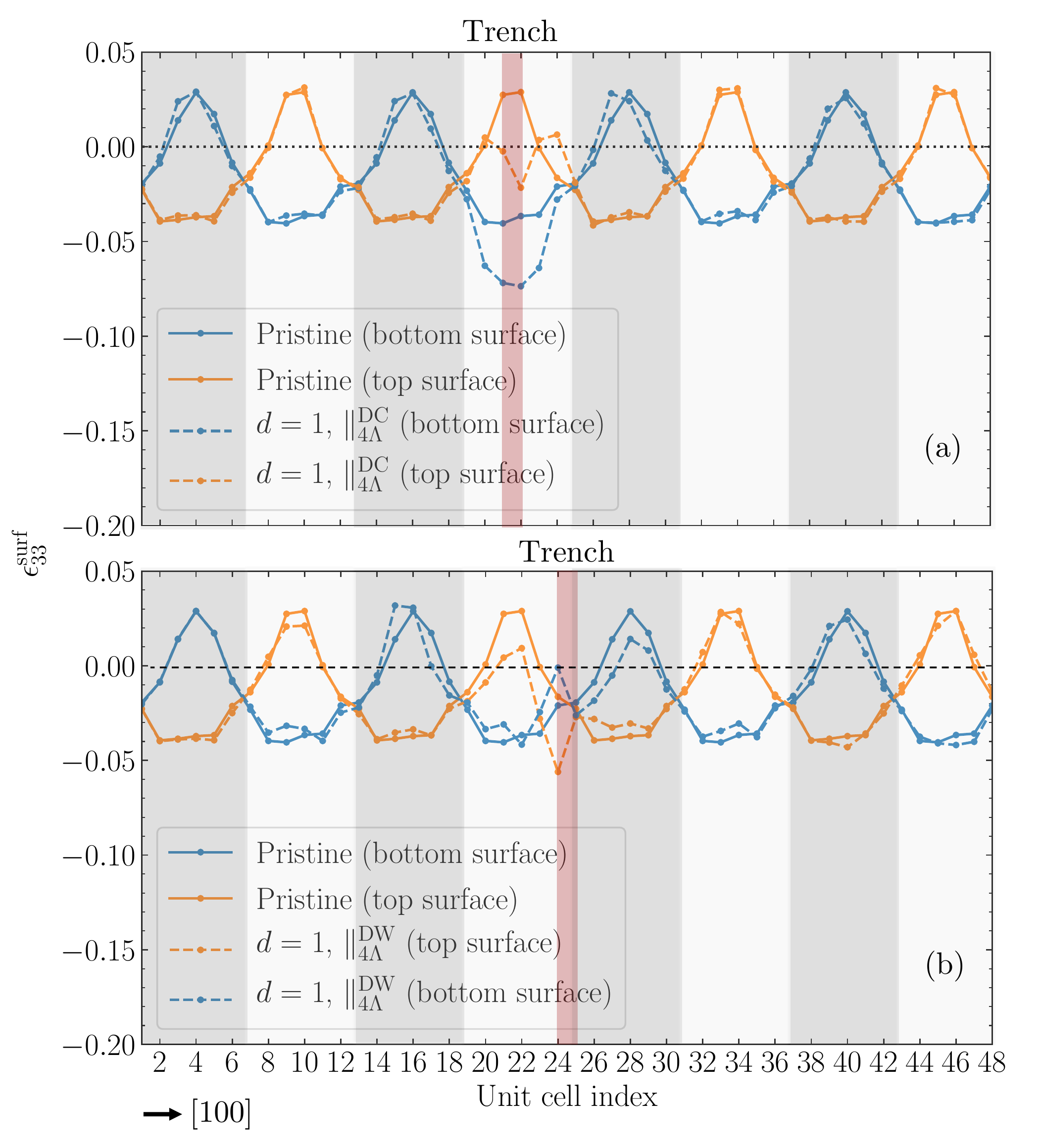}
        \caption{Out-of-plane surface strain $\epsilon_{33}^{\text{surf}}$ along [100] for: a) the $d=1$, $\parallel^{\text{DC}}_{4 \Lambda}$ film and b) the $d=1$, $\parallel^{\text{DW}}_{4 \Lambda}$ film. Pristine film strain is shown for means of comparison. Up domains are colored light grey, down domains are colored a darker grey and the trench site is colored in red.}
        \label{fig:TrenchStrain}
\end{figure}

Motivated by the experimental observation of defect-induced out-of-plane negative strains \cite{Hadjimichael2018}, we examined this effect within our simulations. We define the out-of-plane surface strain $\epsilon_{33}^{\text{surf}}$ relative to the average out-of-plane lattice parameter of the pristine film (4.051 $\text{\AA}$). Negative strains emerge for \textit{all} films within $\approx 4-6$ unit cells of the ST, weakly depending on $d$ apart from $d=3$ which gives rise to large negative strains further from the ST which we postulate is a fictitious artefact of the supercell geometry (the vertical trench-trench interaction is too strong as at $d=3$ they become separated by only a single unit cell). When increasing the ST separation to $4\Lambda$, we observe no change in the range or character of the strain field. These fields are shown in Fig. \hyperref[fig:TrenchStrain]{3}, overlaid with the underlying strain field of the pristine film. For the pristine film, we see that the strain fields of the upper and lower surfaces are a pair of sinusoids in antiphase with asymmetrical antinodes. The origin of this asymmetry is beyond the scope of this letter but will be addressed in a forthcoming publication. We see in Fig. \hyperref[fig:TrenchStrain]{3(a)} that near the ST, the ST-induced negative strain cooperates with the negative strain of the pristine film at the bottom surface (where polarization is directed along [001]) of $\parallel^{\text{DC}}_{4 \Lambda}$. The positive strain of the pristine film is cancelled on the upper surface. Fig. \hyperref[fig:TrenchStrain]{3(b)} shows that $\parallel^{\text{DW}}_{4 \Lambda}$ brings slightly longer-range disorder to the surface strain field than $\parallel^{\text{DW}}_{4 \Lambda}$. While these effects could contribute to the DW alignment mechanism (i.e. piezoelectric and/or flexoelectric contributions), our method does not allow us to separate these contributions from changes in the polarization emerging from the aforementioned depolarizing field minimization. We remark that direct comparison of our results with ref. \cite{Hadjimichael2018} are limited by the drastically different length scales. In \cite{Hadjimichael2018}, defects extend over many domain periods and are less uniform than this study. It is then feasible that different mechanisms could begin to contribute. That being said, in PTO/STO systems, the in-plane axial components have been observed with much longer periods, robust throughout entire samples \cite{Shafer2018}. Then, for larger surface defects, while it may not be possible to discriminate between $\parallel^{\text{DC}}$ and $\parallel^{\text{DW}}$ arrangements, perpendicular-to-wall geometries can be precluded. So long as the defect sufficiently disturbs the axial components, the resulting depolarizing field penalty can be large enough to prevent the condensation of perpendicular-to-wall geometries and thus the ST-DW parallel alignment phenomenon is still observed. \par

In summary, we have investigated the mechanism for experimentally observed parallel DW-ST alignment with state-of-the art DFT simulations using as many as 5,136 atoms. We assert that although the mechanism could have many contributing components, it can be satisfactorily explained by how well the polar texture of a film can adapt to a ST with a given orientation relative to the DW. The preferred orientation features dipole moments in the vicinity of the ST in an arrangement where energy penalties from ST-induced depolarizing fields are most reduced. Subsequently, the preferred polar texture best restores polar continuity and resembles the polar texture of the pristine film. These conditions hold true for the $\parallel^{\text{DC}}$ films which we find to be the most energetically stable arrangement, independent of the presence of AFD modes, larger lateral ST separations and three depths (1, 2 and 3 unit cells). We find large negative strains in the vicinity of our STs for all films. While it is possible that this contributes to the alignment mechanism via piezoelectric and/or flexoelectric effects, we cannot decouple these from other effects within our method. ST-DW parallel alignment is only one consequence of ST engineering. Other exotic polar textures could be engineered in a wide range of ferroelectric nanostructures as demonstrated by the appearance of ST-induced polar cycloidal modulations in $\perp$ + AFD films. Finally, we remark that the system sizes treated within this letter greatly surpass standard \textit{ab initio} studies. This demonstrates the power of $\mathcal{O}(N)$ DFT methods \cite{Li1993, Palser1998, Bowler1999} which we expect to become widely adopted.

\section*{Acknowledgements}
We are grateful for use of the ARCHER/ARCHER2 UK National Supercomputing Service funded by the UKCP consortium EPSRC Grant Ref. No. EP/P022561/1. The authors also thank Pavlo Zubko and Marios Hadjimichael for their careful review this letter.

\nocite{*}

\bibliography{ms.bib}

\end{document}


\author{Jack S. Baker$^{1, 2}$, \& David R. Bowler$^{1, 2, 3}$  \\\small{$^1$\textit{London Centre for Nanotechnology, UCL, 17-19 Gordon St, London WC1H 0AH, UK}} \\ \small{$^2$\textit{Department of Physics \& Astronomy, UCL, Gower St, London WC1E 6BT, UK}} \\ \small{$^3$\textit{International Centre for Materials Nanoarchitectonics (MANA)}} \\ \small{\textit{National Institute for Materials Science (NIMS), 1-1 Namiki, Tsukuba, Ibaraki 305-0044, Japan}}}

\title{\textbf{Supplemental Material} \\ \medskip \Large{Origin of ferroelectric domain wall alignment with surface trenches in ultrathin films}}
\maketitle

\medskip


This document contains supplemental material for the letter ``\textit{Origin of ferroelectric domain wall alignment with surface trenches in ultrathin films}". In Section \ref{simdetail} we provide the full details of the simulation method with careful attention to convergence parameters unique to the $\mathcal{O}(N)$ density functional theory (DFT) calculations used in the main text. We also include a discussion of the accuracy of the method. In Section \ref{supercell} we give finer details related to how PbTiO$_3$ (PTO) films were initialized before structural relaxation as well as justifications for these choices. Further tabulations and example input (including basis sets, pseudopotentials and structure files) relating to the letter are available in \cite{RawData}. Should the reader wish to find further details about the \textsc{CONQUEST} code we refer the reader to a recently written a review article \cite{Nakata2020Large} and to \cite{CQRelease2020} for a public release of the code under an MIT license. If any further information or raw data are required, these can be disseminated following email contact\footnote[2]{Jack S. Baker: \textcolor{blue}{jack.baker.16@ucl.ac.uk}, David R. Bowler: \textcolor{blue}{david.bowler@ucl.ac.uk}}.

\section{Full simulation details \label{simdetail}}

Simulations within the main text use the $\mathcal{O}(N)$ algorithm implemented within the \textsc{CONQUEST} code \cite{Bowler2006, Nakata2020Large} (\text{v1.0.5}). The local density approximation of Perdew \& Wang (LDA-PW) \cite{Perdew1992} is used to represent exchange \& correlation interactions. This choice is motivated by a recent study \cite{Zhang2017} demonstrating that the LDA offers similar levels of accuracy compared with higher rung functionals for describing some important properties. While hybrid and meta-GGA functionals better predict the experimental structural properties, they overestimate the bulk polarization of PTO. Since this is the primary order parameter of the study, the LDA is a good choice as the magnitude of the overestimate is much less. Optimised Vanderbilt norm-conserving pseudopotentials are used to replace core electrons \cite{Hamann2013, Vanderbilt1990}. We use the scalar-relativistic input files provided in the \texttt{PseudoDojo} library (\texttt{v0.4}) \cite{vanSetten2018} which we pass to the \texttt{ONCVPSP} code (\texttt{v3.3.1}) \cite{Hamann2013}. These pseudopotentials treat the Pb 5d 6s 6p, Sr 4s 4p 5s, Ti 3s 3p 4s 3d and the O 2s 2p states as valence, respectively. While many real-space integrals are performed using intuitive analytic operations \cite{Torralba2008}, some are performed on a fine, regular integration grid equivalent to a plane wave cut off of 300 Ha. \par

We use a single-$\zeta$ plus polarisation (SZP) basis set of pseudo-atomic orbitals (PAOs) to describe the above valence states as well as including Pb 6d, Sr 4d, Ti 4p and O 3d polarisation orbitals aimed at increasing the angular flexibility of the basis. The exact details of the PTO basis sets used are described in the supplementary information of \cite{Baker2020_elec} and convergence studies of PAO basis set size for structural and electronic properties are treated exhaustively in \cite{Bowler2019} and \cite{Baker2020_elec} respectively. We also note that the basis set used in this work shares identical generation parameters with the double-$\zeta$ plus double polarization (DZDP) basis sets used in \cite{Baker2020_polar} but with the second $\zeta$'s removed. \par

The $\mathcal{O}(N)$ solver in \textsc{CONQUEST} uses a two-part scheme which hybridizes the canonical purification (CP) strategy in \cite{Palser1998} and the density matrix minimization (DMM) in \cite{Li1993}. A full discussion of how these two parts bind together is presented in \cite{bowler1999} while a comprehensive summary of how DFT calculations can achieve $\mathcal{O}(N)$ scaling can be found in \cite{bowler2012methods} and \cite{Goedecker1999}. We now recommend that readers consult \cite{Nakata2020Large} for definitions for the matrices mentioned in the proceeding discussion. Both the CP and DMM stages of the $\mathcal{O}(N)$ solver require an inverted overlap matrix, $\mathbf{S}^{-1}$. This is completed with Hotelling's method \cite{Hotelling1943} where we truncate the range of $\mathbf{S}^{-1}$ to 6.35 $\text{\AA}$. This choice inverts $\mathbf{S}$ to a tight tolerance of $1 \times 10 ^{-5}$. For the DMM stage \cite{Li1993}, the minimisation tolerance is set to $1 \times 10 ^{-6}$. Self consistency is achieved in tandem with with the DMM stage via a scheme which we call ``mixed-$\mathbf{L}$-SCF". In this scheme, the $\mathbf{K}$-matrix is optimised  with Pulay residual minimisation scheme-direct inversion of the iterative subspace (RMM-DIIS) \cite{Pulay1980} at each DMM step. This is used to update the charge density and the Kohn-Sham potential. We see then that charge density mixing is completed implicitly through the update of $\mathbf{K}$. \par

\begin{figure}
    \centering
    \includegraphics[width=\linewidth]{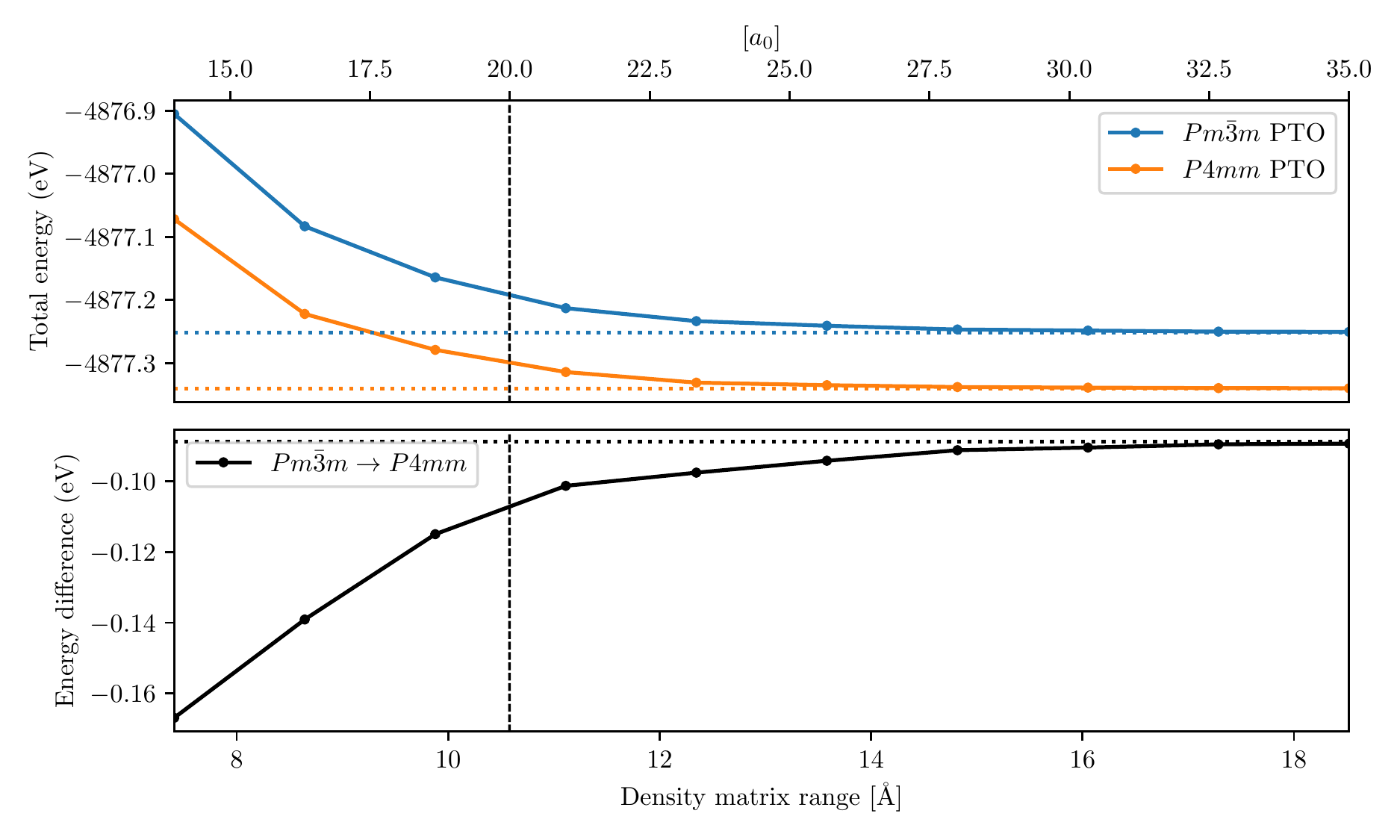}
    \caption{The total energy as a function of density matrix range for the bulk $Pm\bar{3}m$ and $P4mm$ phases of PTO (upper) and the energy difference between them. The vertical black dotted line indicates the density matrix range (10.58 $\text{\AA}$ or 20 $a_0$) used for simulations in the main text.}
    \label{fig:OrderNConverge}
\end{figure}

An important parameter for monitoring the convergence of $\mathcal{O}(N)$ calculations in \textsc{CONQUEST} is the range of the density matrix. Accordingly, we calculate the total energy of the $Pm\bar{3}m$ and $P4mm$ phases as a function of this range. The results can be seen in Figure \ref{fig:OrderNConverge}. To achieve a balance between efficiency and accuracy, we truncate the density matrix range at 10.58 $\text{\AA}$ (20 $a_0$). Table \ref{tab:BulkAccuracy} compares the accuracy of our method with other modes of operation within \text{CONQUEST} and plane wave calculations in \texttt{ABINIT} \cite{Gonze2009, Gonze2016} (\texttt{v8.10.3}) At the exact diagonalisation level (ED, the total energy is yielded from diagonalizing the Hamiltonian matrix), DZDP basis sets return near-plane wave accuracy for lattice constants and the $Pm\bar{3}m$-$P4mm$ energy difference. While the accuracy in the energy difference degrades when the basis set size decreases to SZP and we use the linear scaling (LS) solution (using the aforementioned parameters), we see that the lattice constants and $c/a$ ratios change only marginally. It can then be said that our method overestimates the stability of the FE $P4mm$ phase compared to non-polar phases but retains structural accuracy. Since we only compare polar phases in the main text, we do not expect this to impact our results. \par

To ensure that the above behaviour of our LS calculations are reproducible in other scenarios, we examined the hierarchy of energetic stability for three thin film configurations (with geometries described in \cite{Baker2020_polar} but \textit{without} explicit substrate and initialized with optimal LS unit cells) for the 7 unit cell thick films treated in the main text. In agreement with the higher accuracy method used in \cite{Baker2020_polar}, the polydomain FE flux-closure films were the most energetically stable, followed by monodomain in-plane polarisation ($\mathbf{P} \parallel$ [100]) and then non-polar paraelectric films. In line with the previous paragraph, we found that although the hierarchy of phases is correct, the relative energies of the polar phases compared to non-polar phases were exaggerated by an amount (per atom) comparable to Table \ref{tab:BulkAccuracy}. We proceed then, confident that our method produces high accuracy structures and a correct hierarchy of phases, but, we remark that the exact magnitude of the energy differences are likely exaggerated. We note the this discussion in the main text requires only a correct hierarchy of phases. We reiterate that this exaggeration is only present when comparing polar and non-polar phases. Such comparisons are \textbf{not} performed in the main text.

\section{Supercell initialization \label{supercell}}

While details of the trench configurations are given in the main text, we provide here information related to the polydomain films used as a starting point to insert these trenches. Simulations are performed using a free standing film geometry of a 7 unit cell thick PTO film initialized in a FE flux closure domain configuration. Although no explicit SrTiO$_3$ (STO) substrate is present, we still choose to impose an in-plane strain equivalent to the experimental mismatch of the PTO and STO lattice constants (1.2\%). This strains the system at a level coincident with the experimental observations of domain wall alignment \cite{Thompson2008}. At the PTO unit cell level, this sets the in-plane constants ($a$ and $b$) of the $P4mm$ phase as $a_{\text{STO}} = 3.848$ $\text{\AA}$. Should we hold these constant and relax the c-axis axis and the atomic coordinates in the bulk, we obtain a PTO cell with a tetragonality of 1.077. A linear estimate of the polarisation (using the linear approximation of Resta \cite{Resta1993}) is then 102.81 $\mu C/cm^2$, increased from 87.75 $\mu C/cm^{2}$ in the unstrained case. As described in \cite{Resta1993}, calculating polarization in this way requires knowledge of the Born effective charge tensors. These can be found the supplement of \cite{Baker2020_polar}. We use the strained unit cell to initialise the polydomain structure with equally sized up and down domains separated by a 180$^{\circ}$ domain wall. \par

\begin{figure}
    \centering
    \includegraphics[width=\linewidth]{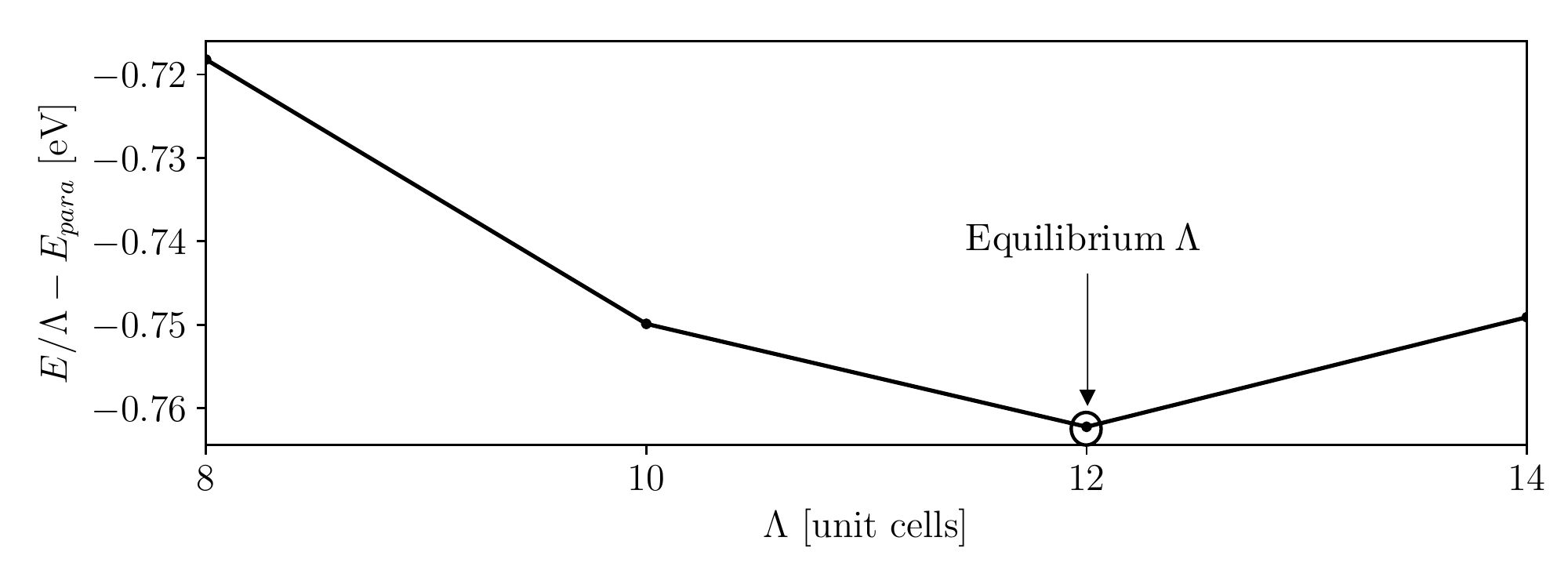}
    \caption{The energy difference (normalized by $\Lambda$) between free standing 7 unit cell thick polydomain films and the paraelectric film as a function of domain period $\Lambda$. The equilibrium period is found at $\Lambda = 12$.}
    \label{fig:FreestandLambda}
\end{figure}

\begin{figure}
    \centering
    \includegraphics[width=\linewidth]{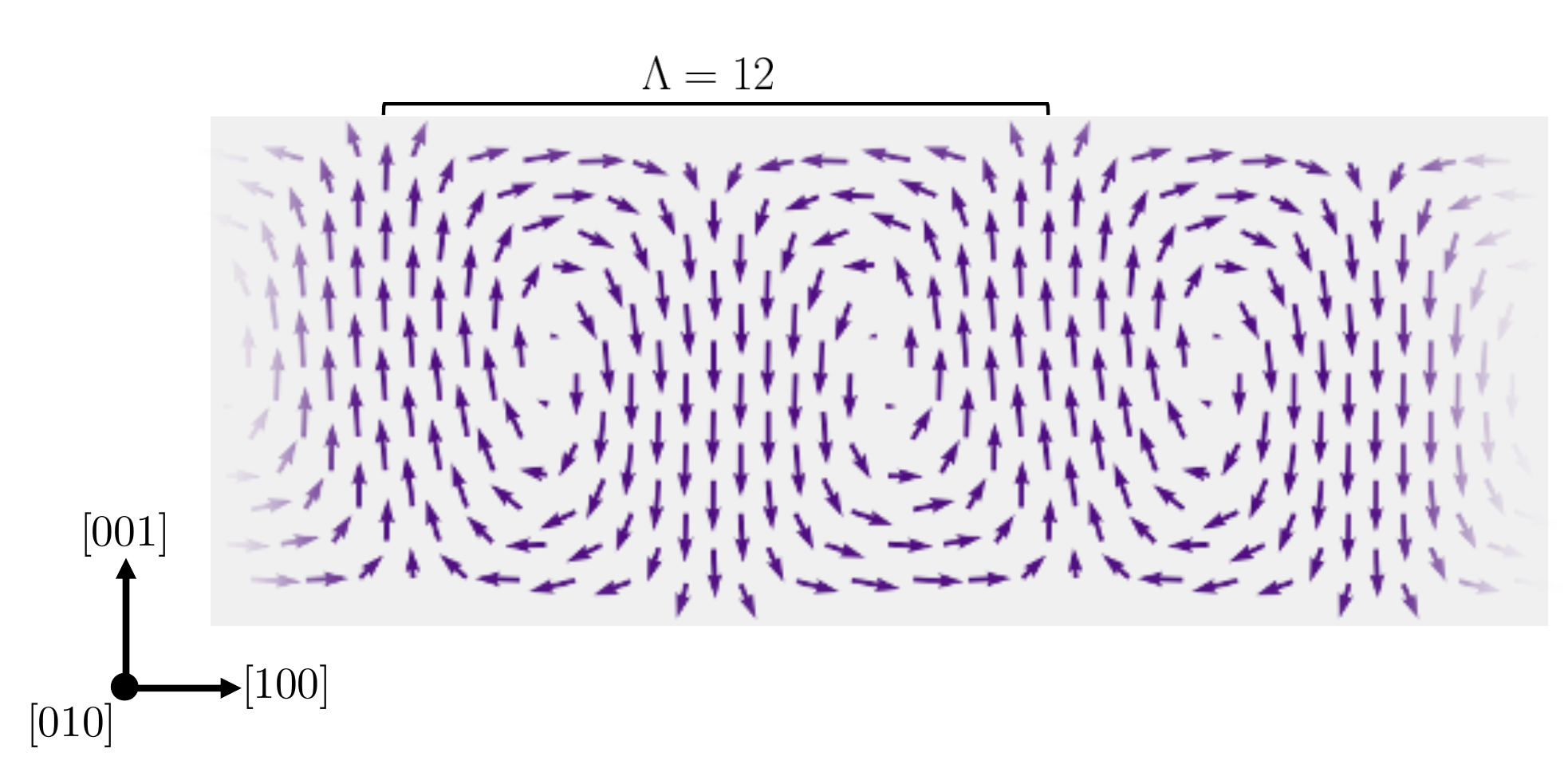}
    \caption{The local polarisation vector field of the 7 unit cell thick polydomain film at the equilibrium domain period $\Lambda = 12$.}
    \label{fig:FreeStandVectorField}
\end{figure}

 In order for our simulations to best reproduce experiment, we choose to work at exactly the theoretical equilibrium domain period. To find this, we prepare four PbO terminated films with $\Lambda = 8, 10, 12$ and 14 where the domain wall is centered on the PbO plane and a vacuum space of 20 $\text{\AA}$ is included. The choice of domain wall centering is largely arbitrary as we do not expect our method to resolve the small energy differences between PbO and TiO$_2$ centering noted in other studies \cite{AguadoPuente2012, Baker2020_polar}. The PbO termination is chosen to invoke the antiferrodistortive (AFD) c$(2 \times 2)$ surface reconstruction \cite{Munkholm2001} which allows us to determine the impact of AFD modes in the main text. Each of these films are relaxed and compared to the energy of the paraelectric films. The results are shown in Figure \ref{fig:FreestandLambda}. We see that the equilibrium $\Lambda$ is 12 unit cells, closely matching the experimentally observed period \cite{Streiffer2002}. Like another work studying free standing polydomain films \cite{Shimada2010Polydomain}, we see the emergence of the flux-closure domain morphology. We show the local polarization field of this relaxed structure in Figure \ref{fig:FreeStandVectorField} (once again calculated using the linear approximation of Resta \cite{Resta1993}). For films including AFD modes, we double the in-plane ([010]) supercell dimensions and introduce small $a^0a^0c^-$ rotations ($\approx 5^{\circ}$ \cite{Baker2020_polar}). We then re-relax the structure. This film is used to initialize the + AFD trench configurations in the main text.


\bibliographystyle{ieeetr}
\bibliography{supplement.bib}


\begin{table}[]
\centering
\caption{The cubic $Pm\bar{3}m$ lattice constant, tetragonality ($c/a$) of the ferroelectric $P4mm$ phase and energy difference between $Pm\bar{3}m$ and $P4mm$ bulk phases of PTO for different levels of accuracy described in the text. These phases are also used in the total energy calculations of Figure \ref{fig:OrderNConverge}.} \medskip
\label{tab:BulkAccuracy}
\begin{tabular}{@{}cccc@{}}
\toprule \toprule
          & $a$, $Pm\bar{3}m$ [\AA] & $c/a$, $P4mm$ & $\Delta E = E_{P4mm} - E_{Pm\bar{3}m}$ [meV/FU] \\ \midrule
PW        & 3.885                   & 1.038         & -48.12                                          \\
DZDP (ED) & 3.905                   & 1.040         & -47.91                                          \\
SZP (ED)  & 3.909                   & 1.043         & -88.86                                          \\
SZP (LS)  & 3.910                   & 1.046         & -106.89                                         \\ \bottomrule \bottomrule
\end{tabular}
\end{table}

